\documentclass[twocolumn,showpacs,preprintnumbers,amsmath,amssymb]{revtex4}
\usepackage{epsf}
\begin{document}

\title{Quasiperiodically driven ratchets for cold atoms}

\author{R.~Gommers$^{1}$, S.~Denisov$^{2}$ and F.~Renzoni$^{1}$}

\affiliation{$^1$Departement of Physics and Astronomy, University College 
London, Gower Street, London WC1E 6BT, United Kingdom\\
$^2$Max-Planck-Institut f\"ur Physik komplexer Systeme,
N\"othnitzer Str. 38, 01187 Dresden, Germany}

\date{\today}

\begin{abstract}
We investigate experimentally the route to quasiperiodicity in a driven ratchet 
for cold atoms, and examine the relationship between symmetries and transport
while approaching the quasiperiodic limit. Depending on the specific form 
of driving, quasiperiodicity results in the complete suppression of transport,
or into the restoration of the symmetries which hold for a periodic driving.
\end{abstract}

\maketitle

The ratchet effect \cite{flashing,doering,mahato}, 
i.e. the possibility of obtaining directed transport of
particles in the absence of a net bias force, has recently been attracting
a considerable interest 
\cite{dykman,flach,super,prost,astumian}. 
Initially introduced to point out the strict limitations on directed transport
at equilibrium imposed by the second principle of thermodynamics
\cite{feynman}, the ratchet effect has subsequently received
much attention as it was identified as a model elucidating the working 
principle of molecular motors \cite{prost}. 
More recently, a considerable activity on ratchets by the condensed 
matter community was stimulated by the possibility of using the ratchet 
phenomenon to realize new types of electron pumps \cite{linke}.

In order to obtain directed transport in the absence of a net bias, the 
ensemble of particles has to be driven out of equilibrium, so to overcome 
the restrictions imposed by the second principle of thermodynamics. 
Additionally, relevant symmetries of the system have to be broken to allow 
directed transport. Theoretical work \cite{flach,super} precisely
identified the relationship between symmetries and transport in the case of 
{\it periodically} driven ratchets, and experiments with cold atoms in optical 
lattices validated the theoretical predictions \cite{schiavoni,gommers}. 
The theoretical analysis was then extended to explore the relationship between 
symmetries and transport for {\it quasiperiodically} driven ratchets, and the 
general symmetries which forbid directed transport were identified \cite{piko,flach04}.

In the present work we investigate experimentally the route to quasiperiodicity
in a driven ratchet for cold atoms, and we examine the relationship between
symmetries and transport while approaching the quasiperiodic limit. It will
be shown that, depending on the specific form of driving, quasiperiodicity
may result in the complete suppression of transport, or into the restoration
of the symmetries which hold for a periodic driving.

Our experiments are based on caesium atoms cooled and trapped in a 
near-resonant driven optical lattice \cite{robi}. The lattice beam geometry
is the same as the one used in our previous experiments \cite{gommers}: 
one  beam (beam 1) 
propagates in the $z$-direction; the three other beams (beams 2--4) 
propagate in the opposite direction, arranged along the edges of a 
triangular pyramid having the $z$ direction as axis. We refer to 
Ref.~\cite{gommers} for further details of the set-up,  
and we summarize here only the essential features. The interference 
between the lattice fields creates a periodic and spatially symmetric potential
for the atoms. The interaction with the light also leads to damping of the
atomic motion, and the level of dissipation can be varied by changing the 
lattice parameters (intensity and detuning from atomic resonance). 
An arbitrary, time-dependent, force can be applied onto the atoms by 
appropriately phase modulating the 
lattice beams. More precisely, a phase modulation $\alpha(t)$ of the
lattice beam 1 will result in an inertial force in the reference frame of
the optical lattice of the form $F(t)=-m\ddot{\alpha}/k_z$, where 
$k_z=2\pi/\lambda_z$ and $\lambda_z/2$ is the distance between neighbouring 
minima in the $z$ direction.

In order to study the relationship between symmetries and transport in the
quasiperiodic limit, we consider a multi-frequency driving, obtained by 
combining signals at three different frequencies: $\omega_1$, $2\omega_1$
and $\omega_2$. For $\omega_2/\omega_1$ irrational, the driving is 
quasiperiodic.  Clearly, in a real experiment $\omega_2/\omega_1$ is 
always a rational number, which can be written as $\omega_2/\omega_1=p/q$,
with $p$, $q$ two co-prime positive integers. However, as the duration of the
experiment is finite, by choosing $p$ and $q$ sufficiently large it is 
possible to obtain a driving which is effectively quasiperiodic on the 
time scale of the experiment. In the present investigation, we consider
two different types of driving forces. We analyze the two cases separately,
as their analysis requires different symmetry considerations.

In the first examined case, we considered a driving of the form:
\begin{equation}
F(t)= A\cos(\omega_1 t) + B\cos(2\omega_1 t+\phi)+C\cos(\omega_2 t+\delta).
\label{eq_driving1}
\end{equation}
Such a force can be applied by phase-modulating in an
appropriate way one of the lattice beams. In the experiment we modulate 
directly the derivative of the phase, i.e. we apply a frequency modulation of 
the form
\begin{equation}
\dot{\alpha}(t)= a\sin(\omega_1 t) + b\sin(2\omega_1 t+\phi)+
c\sin(\omega_2 t+\delta)
\label{eq_driving2}
\end{equation}
which corresponds to a force of the form of Eq.~\ref{eq_driving1} with 
$A=-ma\omega_1/k_z$, $B=-2mb\omega_1/k_z$, $C=-mc\omega_2/k_z$.

We consider first the case of truly periodic driving, i.e. a driving which is 
periodic on the time scale of our experiment. For the current choice of the
experimental parameters the atomic motion is weakly damped. We can therefore
use the symmetry arguments valid in the dissipationless limit, and then include
the corrections due to the damping. For a spatially symmetric potential, and 
a driving $F(t)$ of period $T$  there are two symmetries which need 
to be considered to understand the atomic transport. If $F(t)$ is shift symmetric,
i.e. $F(t)=-F(t+T/2)$, the symmetry $S_a:~(x,t)\to (-x, t+T/2)$
is realized. Furthermore, if $F$ is symmetric, i.e. $F(t)=F(-t)$, the symmetry
$S_b:~(x,t)\to (x, -t)$ is realized. Whenever $S_a$ and/or $S_b$ are
realized, directed transport is forbidden.

\begin{figure}[hbt]
\begin{center}
\mbox{\epsfxsize 3.in \epsfbox{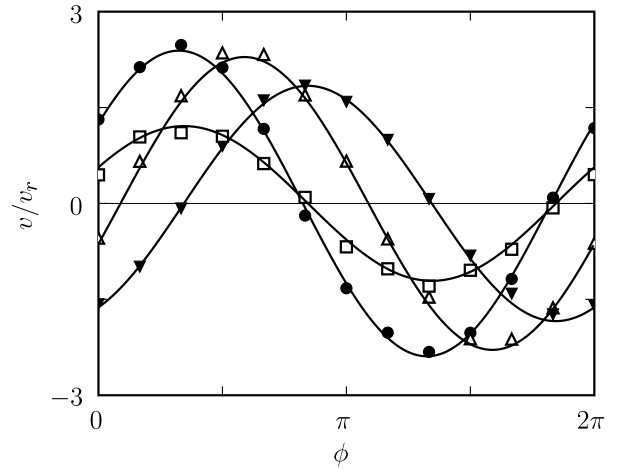}}
\end{center}
\caption{
Experimental results for the average current of atoms in an optical lattice
driven by a force of the form of Eq.~(\protect\ref{eq_driving1}). The average
atomic velocity $v$, rescaled by the recoil velocity $v_r$, is plotted as a
function of the phase $\phi$.  The parameters of the optical lattice
are: the vibrational frequency at the bottom of the well is
$\omega_v = 2\pi 170$ kHz, the laser detuning $\Delta=-24 \Gamma$, where
$\Gamma$ is the excited state linewidth. The parameters of the driving common
to all data sets are: $\omega_1=\omega_2=2\pi 100$ kHz, $a = b=75$ kHz.
The data set with squares corresponds to a simple bi-harmonic driving at
frequencies $\omega_1$, $2\omega_1$, i.e. a driving of the form
of Eq.~(\protect\ref{eq_driving1}) with $C=0$. The other data sets correspond
to a driving including all three harmonics, with $c=75$ kHz, and differ for
the value of the phase $\delta$. Circles corresponds to $\delta=0$, open
triangles to $\delta=\pi/4$ and closed triangles to $\delta=\pi/2$.
The lines represent the best fit of the data with the function
$v=v_{max}\sin (\phi+\phi_0)$.}
\label{fig1}
\end{figure}

As starting point, we examine the case of bi-harmonic driving at frequencies $\omega_1$, $2\omega_1$
($C=0$ in Eq.~\ref{eq_driving1}). The driving breaks the shift symmetry for any 
value of $\phi$, and is time symmetric for $\phi=n\pi$, with $n$ integer. It follows
that directed motion is forbidden for $\phi=n\pi$, and indeed previous work 
\cite{flach,schiavoni} showed
that the current $I$ is of the form $I=I_0\sin\phi$, with $I_0$ a constant.
Now we include the effect of dissipation. Damping breaks the time-reversal
symmetry, and results in an additional 
phase shift $\phi_0$ of the current as a function of the phase $\phi$ \cite{flach,gommers}.
The current 
is in this case $I=I_0\sin (\phi+\phi_0)$. Our experimental results
for the atomic current as a function of the phase $\phi$ are 
shown in Fig.~\ref{fig1}. The squares represent the case discussed above of a
biharmonic drive. These data show the expected
dependence of the current on the phase $\phi$, and will serve for
reference for the rest of the investigation.

We now introduce a third driving at frequency $\omega_2=(p/q) \omega_1$ and
phase $\delta$, see Eq. \ref{eq_driving1}. 
In the case of periodic driving, the same symmetry considerations used
above for the bi-harmonic drive apply. For $\delta=0$ the driving is 
invariant under time reversal, and this forbids directed transport for 
$\phi=n\pi$. Instead, for $\delta\neq 0$ the symmetry under time-reversal
is broken and transport is allowed also for $\phi=n\pi$. In other words, for 
$\delta\neq 0$ the third driving leads to an additional shift of the current
as a function of $\phi$. The magnitude of such a shift depends obviously on
the phase $\delta$, which controls the time-symmetry of the Hamiltonian at
$\phi=n\pi$.
To verify this behaviour, valid for a {\it periodic} driving, we considered
the simple situation with the frequency $\omega_2$ of the third driving 
equal to $\omega_1$ ($\omega_2=\omega_1$, i.e. $p=q=1$) and study the current
as a function of $\phi$, for different values of $\delta$. The experimental
data, shown in Fig.~\ref{fig1}, confirm the behaviour predicted on the 
basis of symmetry considerarions. For $\delta=0$ the third driving does not
introduce any additional phase shift with respect to the case of bi-harmonic 
drive. Instead, the third driving with $\delta\neq 0$ leads to an additional
phase shift of the current as a function of $\phi$, and such a phase shift is 
an increasing function of $\delta$.

\vspace{0.5cm}

We consider now the quasiperiodic limit, which corresponds to an irrational 
ratio of frequencies $\omega_2/\omega_1$. Theoretical work showed that in order
to analyze the relationship between symmetry and transport in the quasiperiodic
case, the two phases $\psi_1\equiv\omega_1 t$
and $\psi_2\equiv\omega_2 t$ can be treated as {\it independent} 
variables.  The symmetries valid in the periodic case can then be generalised 
to the quasiperiodic case. The driving force $F(t)$ is said to be shift-symmetric 
if it changes sign under one of the three transformations 
$\psi_{\alpha}\to\psi_{\alpha}+\pi$ where $\alpha$ is any subset of $\{1,2\}$,
i.e. the $\pi$-shift is applied either to any of the two variables, or to both
of them. If $F$ is shift symmetric, then the system is invariant under the transformation
\begin{equation}
\tilde{S}_a:~x\to -x,~\psi_{\alpha}\to\psi_{\alpha}+\pi
\end{equation}
and directed motion is forbidden \cite{piko,flach04}. The symmetry for time reversal 
is generalised in the same way. The driving is said to be symmetric if 
$F(-\psi_1+\chi_1,-\psi_2+\chi_2)=F(\psi_1,\psi_2)$, with $\chi_1$, $\chi_2$ 
appropriately chosen constants. If the driving is symmetric, in the dissipationless
limit the system is invariant under the transformation
\begin{equation}
\tilde{S}_b:~x\to x,~\psi_j\to -\psi_j+\chi_j~ (j=1,2)
\label{stildeb}
\end{equation}
and directed transport is forbidden \cite{flach04}. These two symmetries determine 
the general transport properties in the quasiperiodic limit. For our driving of 
the form of Eq.~\ref{eq_driving1}, the shift symmetry is broken for any choice of
$\phi$ and $\delta$. The transport is then controlled by the time-reversal
symmetry $\tilde{S}_b$, Eq.~\ref{stildeb}. We notice that the driving is invariant
under the transformation $\psi_2\to -\psi_2+\chi_2$ for any $\delta$, as $\delta$
can be reabsorbed in $\chi_2$. Therefore the invariance under the transformation 
$\tilde{S}_b$ is entirely determined by the invariance of $F$ under the transformation
$\psi_1\to -\psi_1+\chi_1$, i.e. we recover the result for bi-harmonic driving:
$\tilde{S}_b$ is a symmetry, and therefore directed motion is forbidden, for 
$\phi=n\pi$. In other words, in the quasiperiodic limit the additional driving
$C\cos(\omega_2t+\delta)$ does not change the symmetries corresponding to a 
pure bi-harmonic driving, independently of the choice of $\delta$.

\begin{figure}[hbt]
\begin{center}
\mbox{\epsfxsize 3.in \epsfbox{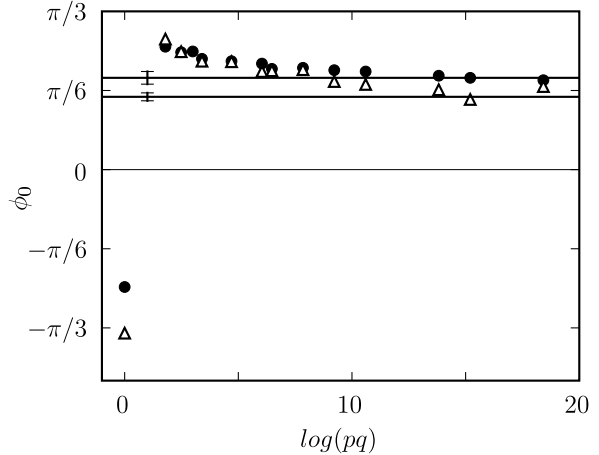}}
\end{center}
\caption{Phase shift $\phi_0$ as a function of the product $pq$ which
characterizes the degree of quasiperiodicity of the driving. The two
data sets correspond to different amplitudes of the driving: $a=150$ kHz
for the circles, and $a = 75$ kHz for the triangles, with $b=c=75$ kHz for
both data sets. The
driving frequencies are: $\omega_1 = 2\pi 100$ kHz, $\omega_2=(p/q)\omega_1$,
and the phase of the driving at frequency $\omega_2$ is $\delta=\pi/2$.
All other parameters are the same as in Fig.~\protect\ref{fig1}. The two
horizontal lines indicate the phase shift $\phi_0$ for bi-harmonic drive,
i.e. in the absence of the driving at frequency $\omega_2$.
}
\label{fig2}
\end{figure}

In order to explore experimentally the route to quasiperiodicity, and investigate the
quasiperiodic limit, we studied the atomic current as a function of $\phi$
for $\omega_2/\omega_1=p/q$, with $p$ and $q$ co-primes. As already discussed,
by increasing $p$ and $q$, the driving will be more and more quasiperiodic
on the finite duration of the experiment. Correspondingly, we will take $pq$
to characterize the degree of quasiperiodicity.
For given $p$ and $q$, we measured the average atomic
velocity as a function of $\phi$. By fitting the data with $v=v_{max}\sin (\phi+\phi_0)$,
we determined the phase shift $\phi_0$, with results as in Fig.~\ref{fig2}.
For small values of the product $pq$, i.e. for {\it periodic} driving,
the third driving at frequency $\omega_2$ leads to a shift $\phi_0$ which 
strongly depends on the actual value of $pq$. For larger values of $pq$, i.e.
approaching quasiperiodicity,  the phase shift $\phi_0$ tends to a constant 
value. Such a value coincides with the phase shift $\phi_0$ measured in the case
of pure bi-harmonic driving (horizontal lines in Fig.~\ref{fig2}), which is determined
by the finite damping of the atomic motion. We also verified that the asymptotic value 
of $\phi_0$ obtained for large $pq$ is independent of the phase difference $\delta$.
These results constitute the experimental proof that in the quasiperiodic limit 
the only relevant symmetries are those determined by the periodic bi-harmonic
driving, and by dissipation. For the specific form of the driving considered, 
quasiperiodicity therefore restores the symmetries which hold in the absence
of the additional driving which produced quasiperiodicity.

\vspace{0.5cm}

We now consider a different driving force, obtained by multiplying the bi-harmonic
driving at frequencies $\omega_1$, $2\omega_1$ with the driving at frequency 
$\omega_2$. This is done by applying a frequency modulation of beam $1$ of the form
\begin{equation}
\dot{\alpha}(t) = c \sin (\omega_2 t+\delta )[ a \sin(\omega_1 t)+
b\sin (2\omega_1 t)]
\end{equation}
which results into a force
\begin{eqnarray}
F(t)&=&-\frac{mc}{k_z}\left\{ \omega_2 \cos(\omega_2 t+\delta)
[ a\sin(\omega_1 t) + b \sin(2\omega_1 t)] \right. \nonumber \\
&+&  \left. \omega_1 \sin(\omega_2 t+\delta)
[a\cos(\omega_1 t) + 2b\cos(2\omega_1 t)] \right\}
\end{eqnarray}
We will show that in this case quasiperiodicity results in the total suppression of
transport. 

\vspace{0.5cm}
\begin{figure}[hbt]
\begin{center}
\mbox{\epsfxsize 3.in \epsfbox{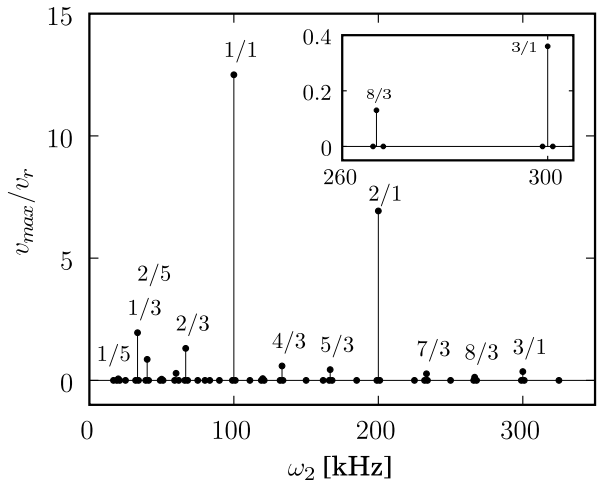}}
\end{center}
\caption{
Maximum average velocity as a function of the driving frequency $\omega_2$.
The data corresponding to a nonzero velocity are labelled by
$p/q=\omega_2/\omega_1$. The inset magnifies a portion of the plot.
}
\label{fig3}
\end{figure}

We examine first the case of periodic driving. We indicate, as before,
$\omega_2=(p/q)\omega_1$. The period $T$ of $F(t)$ is then $T=q T_1=pT_2$,
with $T_i=2\pi/\omega_i$ ($i=1,2$). Under the transformation $t\to t+T/2$ we
have: $\omega_1 t \to \omega_1 t+q \pi$,  $\omega_2 t \to \omega_2 t+p \pi$.
By replacing these transformations in $F(t)$ it is straighforward to see that
$F(t)$ satisfies the shift symmetry $F(t)=-F(t+T/2)$ if $q$ is even, and $p$ is
odd. In this case directed transport is forbidden. If instead this condition is not
satisfied, i.e. if $q$ is odd, directed
transport is not forbidden. In this case directed transport is controlled by
the $S_b$ symmetry which is realized, in the dissipationless limit, if the
driving $F(t)$ is symmetric under time-reversal. The symmetry under time-reversal
depends entirely on the phase $\delta$ of the driving at frequency
$\omega_2$: for $q\delta=(n+1/2)\pi$, with $n$ integer, the driving is symmetric. Otherwise,
the symmetry under time-reversal is broken. The current is
expected to show a sinusoidal dependence on $q\delta-\pi/2$, and dissipation will account
for an additional shift.

In the experiment, we measured the average atomic velocity as a function of
$\delta$ for different values of the driving frequency $\omega_2=(p/q)\omega_1$,
with $p,q$ co-primes. By fitting the data with $v=v_{max}\sin(q\delta+\delta_0)$,
we determined the maximum velocity $v_{max}$ as a function of $\omega_2$. Our
results, shown in Fig. \ref{fig3}, demonstrate the relationship between symmetry
 and transport, valid in the periodic case, discussed above. In fact, a current
is observed only for values of the ratio of driving frequencies
$\omega_2/\omega_1=p/q$ with $q$ odd, which is precisely the  requirement for
the shift symmetry to be broken.

\vspace{0.75cm}
\begin{figure}[hbt]
\begin{center}
\mbox{\epsfxsize 3.in \epsfbox{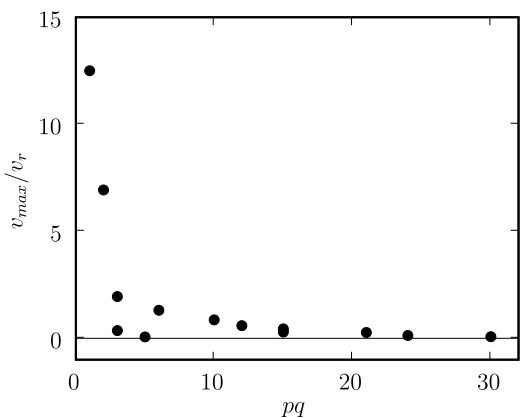}}
\end{center}
\caption{
Maximum average velocity as a function of $pq$, where $p$ and $q$ are
the co-primes defined by the ratio of the driving frequencies:
$p/q=\omega_2/\omega_1$.
}
\label{fig4}
\end{figure}

We turn now to the case of quasiperiodic driving. To analyze
this case, we introduce the two variables $\psi_1=\omega_1 t$ and
$\psi_2=\omega_2 t$, to be treated as independent, and consider the generalized
symmetries $\tilde{S}_a$, $\tilde{S}_b$. It is immediate to verify that
$F$ changes sign under the transformation $\psi_2\to \psi_2+\pi$, i.e.
$F$ is shift symmetric with respect to $\psi_2$. It follows that the
system is invariant under the generalized symmetry $\tilde{S}_a$. Furthermore,
in the quasiperiodic limit the system is also invariant under $\tilde{S}_b$.
Directed transport is therefore forbidden.
In order to study the transition to quasiperiodicity, we re-arrange the
data of Fig.~\ref{fig3} as a function of $pq$ which characterizes the
quasiperiodic caracter of our driving on the finite duration of the experiment.
The results are shown in Fig.~\ref{fig4}. It appears that for large $pq$ values
the amplitude of the atomic current decreases to zero. This demonstrates
that directed transport is destroyed in the quasiperiodic limit, as a results of
the restoration of the shift symmetry of the driving.

In conclusion, in this work we studied experimentally the route to quasiperiodicity 
in a driven ratchet for cold atoms. We examined the relationship between symmetries 
and transport for two different types of driving. Depending on the 
specific form of driving, quasiperiodicity results in the complete suppression
of transport, or into the restoration of the symmetries which hold for a periodic 
driving. Our results also demonstrate that by using a multi-frequency driving 
it is possible to precisely control the directed transport by just varying 
the frequency of one driving. Multi-frequency driving also allows one to implement subtle
mechanisms of control of the current direction, by introducing appropriate 
time-correlations \cite{doering}.

We thank S.~Flach, H.~Maier, W.P.~Schleich and S.~Stenholm for useful discussions.
Financial support from EPSRC, UK and the Royal Society is acknowledged.

\end{document}